\documentclass[twocolumn,showpacs,floats,floatfix,superscriptaddress,aps,pra]{revtex4-1}
\usepackage{amsfonts}
\usepackage{amssymb}
\usepackage{amsmath}
\usepackage{calc}
\usepackage{graphicx}
\usepackage{bm}

\usepackage[normalem]{ulem}
\usepackage{amsmath,amssymb}
\usepackage{multirow}
\usepackage{xcolor,soul}
\usepackage{srcltx}

\begin{document}

\author{Stefano Longhi$^{*}$} 
\affiliation{Dipartimento di Fisica, Politecnico di Milano, Piazza L. da Vinci 32, I-20133 Milano, Italy}
\affiliation{IFISC (UIB-CSIC), Instituto de Fisica Interdisciplinar y Sistemas Complejos, E-07122 Palma, Spain}
\email{stefano.longhi@polimi.it}

\title{Unraveling the Non-Hermitian Skin Effect in Dissipative Systems}

  \normalsize

%\date{.}

%
\bigskip
\begin{abstract}
\noindent  
 The non-Hermitian skin effect, i.e. eigenstate condensation at the edges in lattices with open boundaries, is an exotic manifestation of non-Hermitian systems. 
In Bloch theory, an effective non-Hermitian Hamiltonian is generally used to describe dissipation, which however is not norm-preserving and neglects quantum jumps. Here it is shown that in a self-consistent description of the dissipative dynamics in a one-band lattice, based on the stochastic Schr\"odinger equation or Lindblad master equation with a collective jump operator, the skin effect and its dynamical features are washed out. Nevertheless, both short- and long-time relaxation dynamics provide a hidden signature of the skin effect found in the semiclassical limit. In particular, relaxation toward a maximally mixed state with the largest von Neumann entropy in a lattice with open boundaries is a manifestation of the semiclassical skin effect.
\end{abstract}

\maketitle

{ \it Introduction.}  
Dissipative lattices, where energy or particle number are not conserved, show intriguing
topological properties and phase transitions that are attracting a great interest in different areas of physics \cite{r1,r2,r3,r4,r5,r6,r7,r8,r9,r10,r11,r12,r13,r14,r15,r16,r17,r18,r19,r20,r21,r22,r23,r24,r25,r26,r27,r28,r29,r30,r31,r32,r33,r34,r35,r36,r37,r38,r39,r40,r40b,r41,r42,r43,r44,r45,r46,r47,r48,r49,r50,r51,r52,r53,r53a,r54,r55,r56,r57,
r58,r59,r60,r61,r62,r62b}. 
Band theory describes open systems by non-Hermitian Hamiltonians \cite{r14,r17,r25,r26,r41,r45} and predicts a wealth of exotic features such as  a
strong sensitivity of the energy spectrum on boundary conditions \cite{r11,r12,r15,r17}, the non-Hermitian skin effect (NHSE), i.e. the condensation of bulk modes at the edges
\cite{r13,r14,r15,r16,r17,r18,r45,r53}, and
breakdown of bulk-boundary correspondence based on Bloch topological invariants \cite{r6,r12,r13,r14,r15,r30,r31,r32,r33,r34,r35,r36,r46,r47,r48,r49,r51}. However, on a fundamental level the non-Hermitian description generates a non-unitary dynamics and ignores quantum jumps. The natural description of quantum dynamics in open systems is provided by master equations in Lindblad form or stochastic Schr\"odinger equations, where stochastic terms ensure unitary time evolution. This has motivated the search for a topological classification of dissipative systems and for exotic phenomena like the skin effect beyond non-Hermitian band theory \cite{r24,r54,r59,r60}. Recently, topological classifications of dissipative systems based on the complex spectrum of the Lindbladian superoperator or on quantum jump dynamics \cite{r54,r59} have bee suggested, and the prediction of important phenomena, like the boundary-dependent damping dynamics and the Liouvillian skin effect \cite{r24,r60}, have been disclosed. An open question is whether signatures of the NHSE in non-Bloch band theory persist when considering stochastic terms or quantum jumps, i.e. beyond a mean-field theory.\\
 In this Rapid Communications it is shown that the NHSE of non-Hermitian band theory and its dynamical signatures  are washed out in a minimal model of stochastic dynamics which restores a unitary time evolution. Nevertheless, both short and long-time relaxation dynamics provide hidden signatures of the NHSE. In particular, relaxation toward  a maximally mixed state with the largest von Neumann entropy in a one-band system with open boundaries is the manifestation of the semiclassical NHSE.\\
 \\
{\it Dissipative dynamics and skin effect.} In the framework of single-particle Bloch theory, a one-band dissipative lattice made of $N$ sites under either periodic (PBC) or open (OBC) boundary conditions  is described in Wannier basis $| n \rangle$ by an effective $N \times N$ non-Hermitian matrix Hamiltonian $\hat{H}_{eff}$, which can be written as the sum of Hermitian $\hat{H}$ and anti-Hermitian $i\hat{A}=-(i/2) \hat{P}^2$ parts, i.e.
\begin{equation}
\hat{H}_{eff}=\hat{H}-(i/2) \hat{P}^2
\end{equation}
with $\hat{H}^{\dag}=\hat{H}$ and $\hat{P}^{\dag}=\hat{P}$ for a purely dissipative lattice. Under PBC, in momentum space the Hamiltonian is diagonal and takes the form
\begin{equation}
H_{eff}(k)=H(k)-(i/2) P^2(k)
\end{equation} 
where $k$ is the Bloch wave number and $P(k)$ can be taken a non-negative function. Here we assume that the Hermitian dynamics, described by $\hat{H}$, shows time reversal symmetry so that $H(-k)=H(k)$ in momentum space and $\hat{H}^T=\hat{H}$ in physical space. For any Hermitian Hamiltonian, the bulk energy spectrum in the thermodynamic limit $N \rightarrow \infty$ becomes independent of boundary conditions, and the energy spectra may differ just for the appearance of isolated energies corresponding to edge states under OBC. However, for a generic non-Hermitian Hamiltonian the bulk eigenenergies under PBC and OCB rather generally differ considerably in the thermodynamic limit owing to the NHSE \cite{r13,r14,r15,r16,r17,r18}. In particular,  the energy spectrum of $\hat{H}_{eff}$ is different for PBC and OBC whenever $P(-k) \neq P(k)$,  i.e. the NSHE does not arise provided that the system possesses the additional symmetry $\hat{P}^T=\hat{P}$. In fact, for $P(-k) \neq P(k)$ the PBC energy spectrum describes a closed loop in the complex energy plane with a finite number of self-intersections \cite{r63}, while under OBC the energy spectrum must collapse to one (or a set of) closed curves in the interior of the closed loop \cite{r17,r29,r61}.  As an example, PBC and OBC bulk energy spectra for the dissipative lattice with
\begin{equation}
H(k)= 2 J \cos k+2T \cos(2 k)  \; , \; P(k)= R [1+ \cos (k+ \varphi) ]
\end{equation}
are depicted in Fig.1. Note that a non vanishing phase $\varphi \neq 0$, breaking the symmetry $P(-k)=P(k)$, results in the NHSE.
\begin{figure}[htbp]
  \includegraphics[width=87mm]{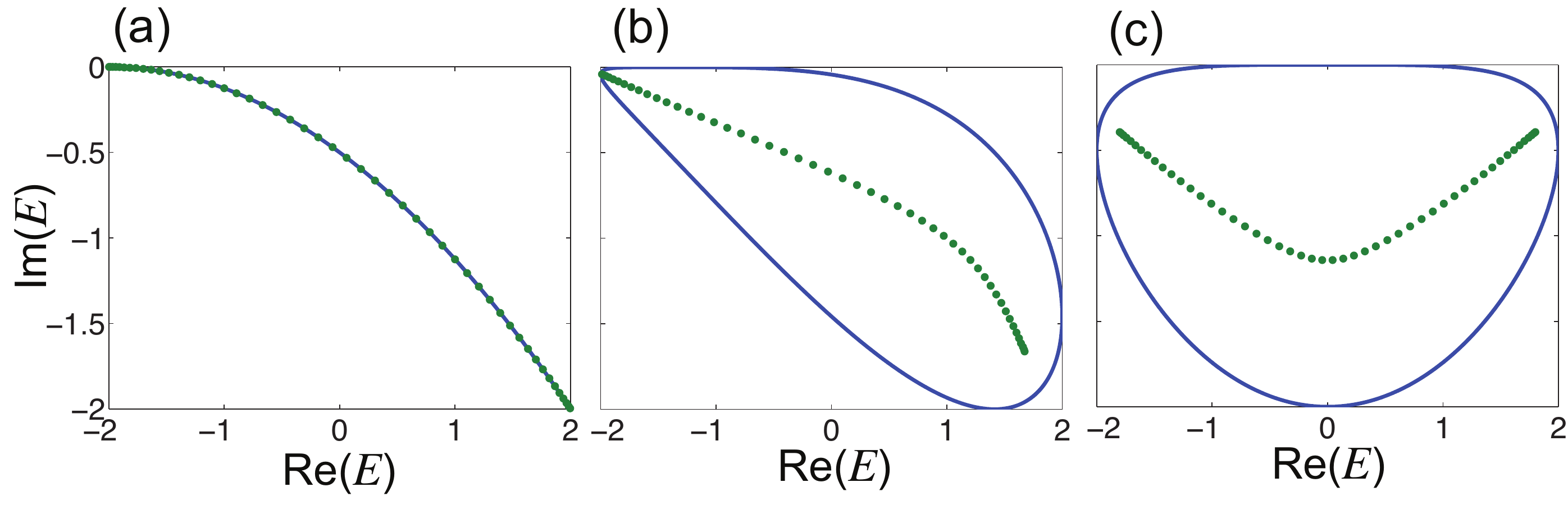}\\
   \caption{(color online) Energy spectrum in complex plane of the non-Hermitian lattice, defined by Eqs.(1) and (3), for $J=R=1$, $T=0$ and for a few increasing values of the phase $\varphi$: (a) $\varphi=0$, (b) $\varphi=\pi/4$, (c) $\varphi=\pi/2$. Solid curves refer to PBC, while circles refer to OBC in a lattice comprising $N=50$ sites. The NHSE appears in (b) and (c), where PBC and OBC energy spectra are distinct.}
\end{figure}
The description of dissipation based on the use of $\hat{H}_{eff}$ in the Schr\"odinger equation suffers from the fact that the state vector $| \psi(t) \rangle$ undergoes a non-unitary evolution and neglects quantum jumps. 
 As such a description  can be regarded as a semiclassical limit of the Markovian dynamics of the open quantum system and can naturally appear in post selection of quantum trajectories \cite{r64,r64b}, an increasing attention is currently devoted to unveil NHSE beyond the semiclassical limit \cite{r24,r54,r59,r60}. A unitary dynamics is restored by considering a {\em stochastic} Schr\"odinger equation, where stochastic terms are added to the deterministic evolution of $| \psi(t) \rangle$ so as to preserve the norm \cite{r65}. This description is equivalent to the use of a master equation in Lindblad form \cite{r65}. As the choice of the stochastic terms (i.e. jump operators) is not unique \cite{r65,r66} and  a detailed microscopic knowledge of the system-bath coupling would be required, here we focus our attention to a model that exploits a single non-local jump operator \cite{r64}, corresponding to the stochastic Schr\"odinger equation (in the It\^{o} interpretation  \cite{r65,r65b})
 \begin{equation}
 i d | \psi(t) \rangle= \hat{H}_{eff} | \psi (t) \rangle dt +  \xi(t) \hat{P} | \psi(t) \rangle  dt
 \end{equation}
 where $\xi(t)$ is a zero-mean delta-correlated white noise, i.e. $\overline{ \xi(t)} =0$ and  $\overline{ \xi(t) \xi(t^{\prime})}=\delta (t-t^{\prime})$. Note that the mean value $\overline {| \psi(t) \rangle}$, averaged over all realizations of noise, evolves as $ \overline{|\psi(t) \rangle} = \exp(-i t \hat{H}_{eff}) \overline{|\psi(0) \rangle} $, corresponding to the mean-field (or semiclassical) dynamics with effective non-Hermitian Hamiltonian $\hat{H}_{eff}$. 
 The associated Lindblad master equation for the density operator $\hat{\rho}= \overline{| \psi(t) \rangle \langle \psi(t)|}$ involves a single jump operator $\hat{P}=\hat{P}^{\dag}$ and reads  \cite{r65,r66}
 \begin{equation}
 \frac{d \hat{\rho}}{dt}= -i [ \hat{H}, \hat{\rho} ] - \frac{1}{2} \left(  \hat{P}^2 \hat{\rho}+  \hat{\rho} \hat{P}^2-2 \hat{P} \hat{\rho} \hat{P} \right) \equiv \mathcal{L} \hat{\rho}
 \end{equation}
 where $\mathcal{L}$ is the Liouvillian superoperator.
  We stress that this model with a collective jump operator may not be sufficient to fully
capture the underlying dissipative process, however it has been suggested as a minimal description (involving a single jump operator) of the open 
system dynamics beyond the semiclassical limit \cite{r64}. Also, the stochastic or master equation approach can describe system dynamics under classical noise \cite{r65,r67,r68}, where the model of Eq.(4)
 can be realized.\\  
  \\
  {\it Relaxation dynamics.}
 The spectrum and corresponding $N^2$ eigenmodes of the Liouvillian superoperator $\mathcal{L}$ are rather generally different for lattices with PBC and OBC, so that relaxation and decoherence dynamics are boundary-dependent \cite{r24}. In particular, the exponential localization of the eigenmodes of  $\mathcal{L}$  at the edges results in so-called Liouvillian skin effect \cite{r60}, which is responsible for slowing down of relaxation processes without gap closing. Here we consider a different scenario, where the semiclassical NHSE is washed out by the stochastic dynamics and $\mathcal{L}$ does not show skin modes. The main result of this work is that the semiclassical NHSE can be nevertheless unraveled by looking at the relaxation dynamics both at short and long time scales. Let us first consider the {\it short-time} (bulk) relaxation dynamics, where initial excitation is spatially confined far from the boundaries and edge effects are negligible. In this case Eq.(4) can be solved in momentum space, where all operators are diagonal. After expanding the state vector in Bloch basis as $|\psi(t) \rangle = \int_{-\pi}^{\pi} dk \Psi(k,t) | k \rangle$, where $k$ is the Bloch wave number and $| k \rangle \equiv(1/ \sqrt{2 \pi}) \sum_n \exp(ikn) | n \rangle$ the Bloch basis, one obtains
\begin{equation}
\Psi(k,t)=\Psi(k,0) \exp[-i H(k)t-iP(k) W(t)]
\end{equation}
where $W(t)$ is a Wiener process with $\overline{W(t)}=0$ and $\overline {W^2(t)}=t$. The corresponding evolution of the density operator in Bloch  basis, $\rho_{k,k^{\prime}}(t)=\langle k | \rho | k^{\prime} \rangle$, reads
\begin{equation}
\rho_{k,k^{\prime}}(t)  =\overline{ \Psi(k,t) \Psi^*(k^{\prime},t)}= \rho_{k,k^{\prime}}(0) \exp[iG(k,k^{\prime})t]
\end{equation}
where we have set $G(k, k^{\prime})=H(k^{\prime})-H(k)+(i/2)[P(k^{\prime})-P(k)]^2$. The signature of the semiclassical NHSE in the short time scale is clearly observed looking at the relaxation dynamics of density matrix elements $\rho_{n,m}(t) = \langle n | \hat{\rho} | m \rangle$ in Wannier basis. In particular, assuming an initial pure state with excitation at site $n=0$, i.e. $\rho_{n,m}(0)= \delta_{n,0} \delta_{m,0}$, one has (see \cite{r66} for details)
 \begin{equation}
\rho_{n,m}(t)=\frac{1}{4 \pi^2} \iint dk dk^{\prime}  \exp [ ikn-ik^{\prime}m+iG(k,k^{\prime})t ].
 \end{equation}
For $P(-k)=P(k)$, i.e. when $\hat{H}_{eff}$ does not show the NHSE, the relaxation process is highly symmetric around $n=m=0$, namely $\rho_{-n,m}=\rho_{n,-m}=\rho_{-n,-m}=\rho_{n,m}$ owing to the even symmetry of the spectral term $G(k,k^{\prime})$ under momentum inversion. This means that coherence is created and maintained during the relaxation process. On the other hand, for $P(-k) \neq P(k)$, i.e. when $\hat{H}_{eff}$ shows the NHSE, using a multivariate saddle-point method \cite{r69} it can be shown \cite{r66} that the slowest decaying terms are those along the main diagonal $n=m$, i.e. $^{\prime}$populations$^{\prime}$, while terms along the anti-diagonal $n=-m$, i.e. $^{\prime}$coherences$^{\prime}$, decay faster. Such a distinct behavior in relaxation dynamics is illustrated in Fig.2 for the model defined by Eq.(3). Note that the ballistic spreading of $^{\prime}$populations$^{\prime}$, i.e. of $\rho_{n,n}(t)$, is symmetric at around $n=0$, while in the semiclassical limit one would expect a {\it unidirectional} flow when $\hat{H}_{eff}$ shows the NHSE \cite{r29}, as illustrated in Fig.6 of \cite{r66}. This means that quantum jumps fully change the spreading features of excitation along the lattice as compared to the mean-field model. Yet, the signature of the semiclassical NHSE is visible looking at the coherences, which decay faster lading to a characteristic {\it elongated} spreading pattern when the systems shows the NHSE in the semiclassical limit.

\begin{figure}[htbp]
  \includegraphics[width=87mm]{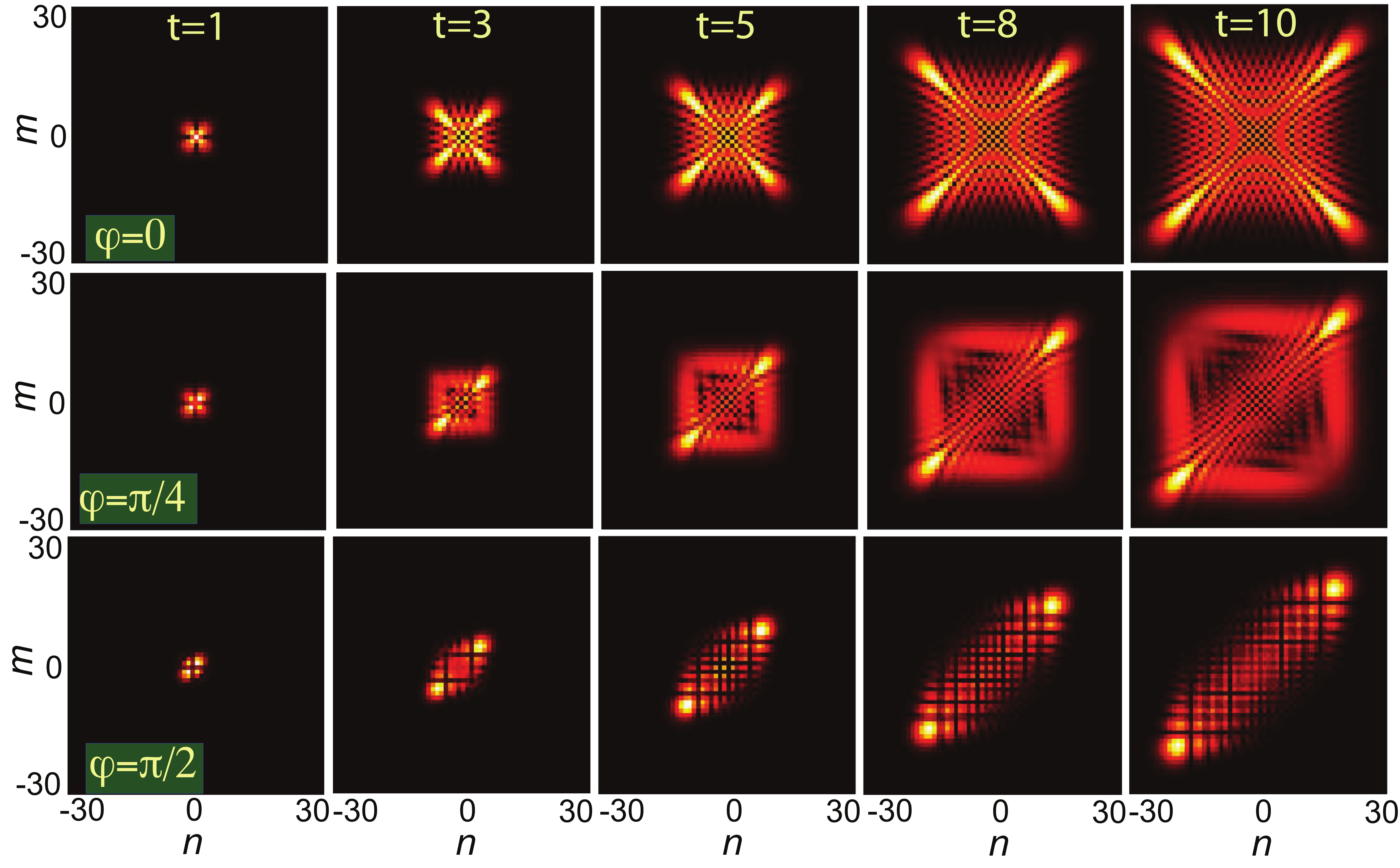}\\
   \caption{(color online) Bulk (short-time) relaxation dynamics in the dissipative lattice defined by Eq.(3) for three values of the phase $\varphi$ ($0$, $ \pi/4$ and $\pi/2$). The panels show snapshots of the density matrix $| \rho_{n,m}(t)|$ in Wannier basis at successive times on a pseudocolor map. The initial state is the pure state $\rho (0) = |0 \rangle \langle 0|$, corresponding to a particle localized at site $n=0$. Lattice parameters are $J=R=1$ and $T=0$.}
\end{figure}

\begin{figure}[htbp]
  \includegraphics[width=87mm]{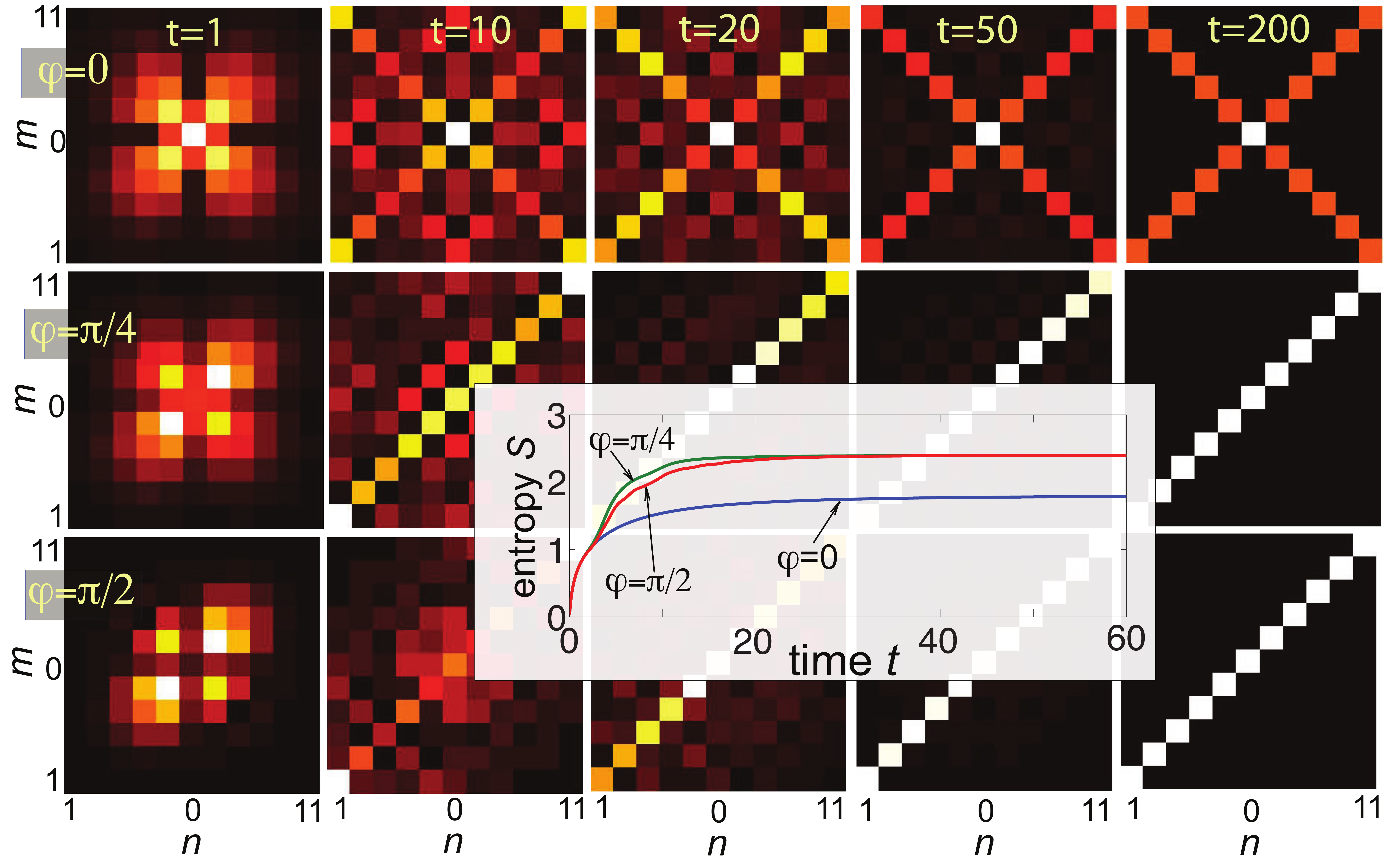}\\
   \caption{(color online) Same as Fg.2, but in a lattice comprising $N=11$ sites with open boundaries. Initial state is $\hat{\rho}(0)=|6 \rangle \langle 6|$.
   Inset: evolution of the von Neumann entropy $S$ for the three phases $\varphi=0$, $\pi/4$, and $\pi/2$. The long-time asymptotic state is a maximally mixed state, corresponding to the largest value of entropy $S={\rm ln} N$, except than $\varphi=0$.}
\end{figure}

The other hidden signature of the semiclassical NHSE is found looking at the long-time relaxation dynamics, where edge effects can not be neglected. In this case the relaxation process is established by the non-decaying eigenvectors of the Liouvillian superoperator. Interestingly, more than one stationary state can exist under certain symmetries of  $\mathcal{L}$ \cite{r70,r71,r72}, which in our model depend on the symmetries of $\hat{H}$ and $\hat{P}$. The $N^2$ eigenvalues $\lambda_l$ of $\mathcal{L}$ always appear in complex conjugate pairs and satisfy the condition ${\rm Re}(\lambda_l) \leq 0$. If $\mathcal{L}$ shows a single non-decaying eigenvector $\hat{\rho}^{(s)}$ with zero eigenvalue, the system relaxes toward the stationary state $\hat{\rho}^{(s)}$. For the master equation (5) it can be readily shown that the state ${\rho}^{(s)}_{n,m}= (1/N) \delta_{n,m}$ is an eigenvector of $\mathcal{L}$ with zero eigenvalue. Such a stationary state corresponds to a maximally mixed state, with von Neumann entropy $S(\hat{\rho})=-{\rm tr} ( \hat{\rho} \log \hat{\rho})$ reaching its largest value $S(\hat{\rho}^{(s)})= \ln N$. In the absence of additional symmetries of $\hat{P}$, namely for $\hat{P}^T \neq \hat{P}$, $\hat{\rho}^{(s)}$ is the only non-decaying eigenvector of $\mathcal{L}$: this means that, if the mean-field dynamics displays the NHSE, then the relaxation dynamics drives the system toward a maximally mixed state. However, when the system shows the additional symmetry $\hat{P}^T=\hat{P}$, corresponding to the absence of the NHSE in the mean-field limit, besides $\hat{\rho}^{(s)}$ there is {\it at least} another stationary state, given by $\hat{\rho}^{(s_a)}_{n,m} =(N+1)^{-1} \left( \delta_{n,m}+\delta_{n,N-n+1} \right)$; technical details are given in \cite{r66}. We also mention that more than two stationary states can arise when $\hat{P}^T=\hat{P}$ and the matrices $\hat{P}$ and $\hat{H}$ are tridiagonal, i.e. when there are only nearest-neighbor hopping [like in model (3) with $T=0$] \cite{r66}. This implies that, when the system possess the symmetry $P(-k)=P(k)$ and the NHSE is prevented in the semiclassical limit, the dissipative process does not drive the system toward a maximally mixed state for rather arbitrary initial conditions, and the von Neumann entropy $S$ remains below the value ${\rm ln} N$, as illustrated in Figs.3 and 4. The long-time evolution of density matrix in a lattice with open boundaries, shown in Fig.3 for the Hamiltonian (3) with $T=0$, clearly  indicates that the system relaxes to a maximally mixed state except than $\varphi=0$. Such a behavior is closely related to the spectrum of the Liouvillian superoprator $\mathcal{L}$, which is shown in Fig.4. While for $\varphi \neq 0$ the Liouvillian ${\mathcal L}$ has a single non decaying stationary state, corresponding to the largest von Neumann entropy $S= {\rm log} N$, for $\varphi=0$ there are $N$ distinct stationary (non-decaying) states \cite{r62}. In the latter case the stationary state in the long-time limit depends on the initial state $\rho_{n,m}(0)$ of the system, and does not correspond rather generally to the maximally mixed state. Numerical diagonalization of $\mathcal{L}$ shows that, while for $\varphi \neq 0$ $\hat{H}_{eff}$ shows the NHSE, the eigenstates of $\mathcal{L}$ {\em are not} skin modes, i.e. we do not have here the Liouvillian skin effect \cite{r60}.  We stress that the presence of the NHSE in the semiclassical description and the relaxation toward  a maximally mixed state in the master equation description have the same physical ground, i.e. breaking of the symmetry $P(-k)=P(k)$. \\ The above results are rather general ones and persist in models with long-range hopping [e.g. $T \neq 0$ in Eq.(3)]. In the Supplementary Material \cite{r66} we discuss another example, the stochastic extension of the Hatano-Nelson model \cite{r1}, which displays a long-range matrix $\hat{P}.$\\  
\begin{figure}[htbp]
  \includegraphics[width=87mm]{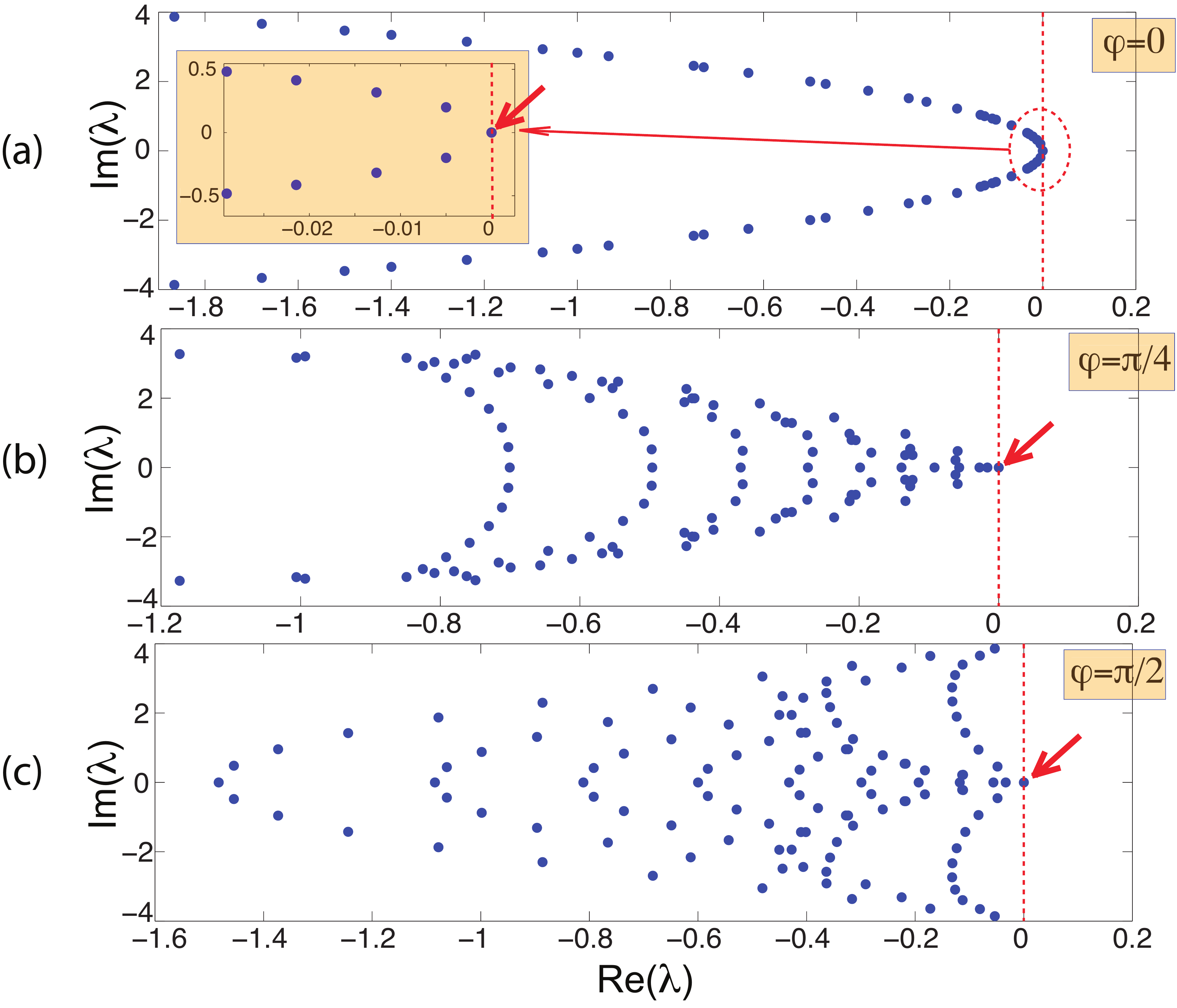}\\
   \caption{(color online) Numerically-computed eigenvalues $\lambda$ of the Liouvillian superoperator $\mathcal{L}$ corresponding to the simulations of Fig.3. The arrows show the zero eigenvalue of $\mathcal{L}$, which is simple in (b) and (c) and $N$-fold degenerate in (a).}
\end{figure}
\begin{figure}[htbp]
  \includegraphics[width=87mm]{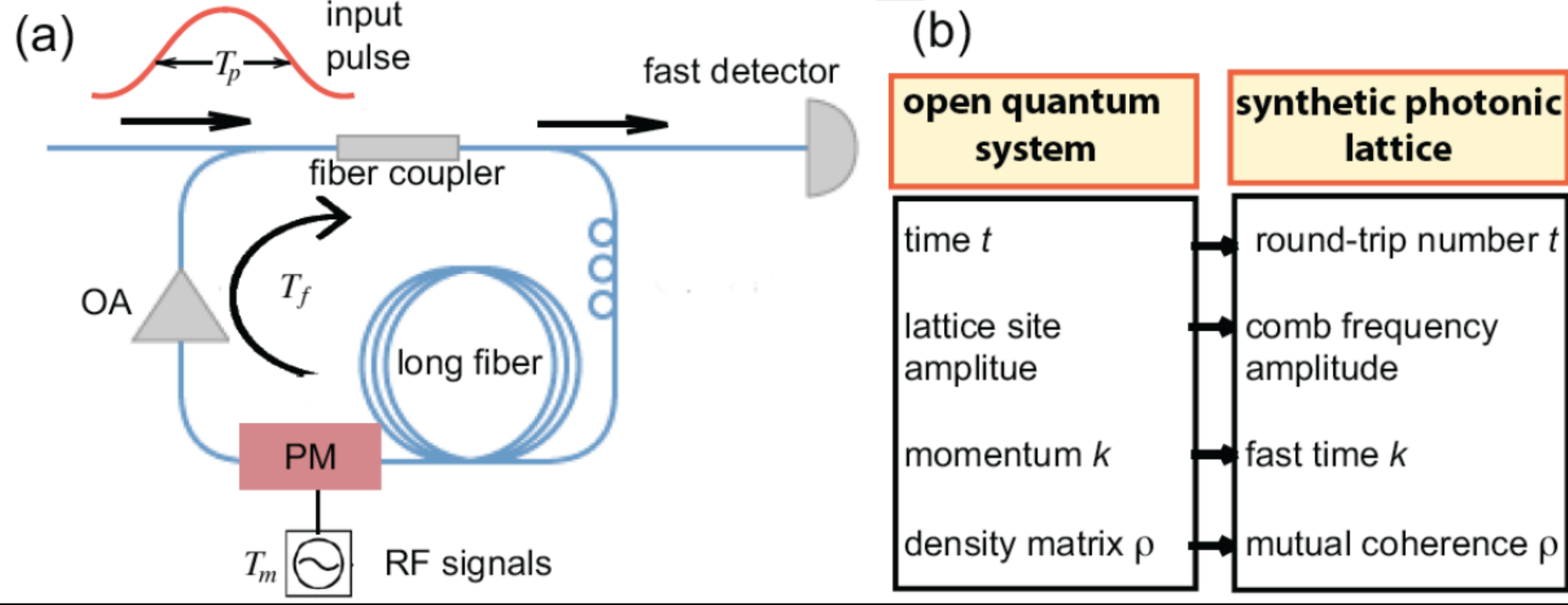}\\
   \caption{(color online) (a) Setup of a spectral photonic lattice. An optical pulse circulating in a fiber loop containing an optical amplifier (OA) and a phase modulator (PM), driven by a period RF signal with stochastic amplitudes at successive transits, realizes the stochastic process (4) in frequency domain. (b) Equivalence between a dissipative quantum lattice and a spectral photonic lattice.}
\end{figure}

{\it Photonic simulation of stochastic dynamics.} Classical systems with unitary dynamics driven by noise can provide a feasible platform to emulate the stochastic Sch\"odinger equation (4) with a collective jump operator. In such classical systems, decoherence arises by averaging over an ensemble
of stochastic but unitary dynamics. A possible implementation in optics is provided by {\em spectral} photonic lattices \cite{r73,r74,r75,r76,r77,r78}. A long light pulse circulating in a fiber loop with negligible dispersion and periodically kicked by a phase modulator realizes a spectral lattice where the spectral curves $H(k)$ and $P(k)$ in Eq.(4) are determined by the waveform that drives the modulator. A schematic of the photonic setup and the equivalence between a quantum dissipative lattice  and a spectral photonic lattice  are illustrated in Fig.5. A long optical pulse with envelope $\Psi(k,0)$  of temporal duration $T_p$ is injected by an optical coupler into a long fiber loop, where $k$ denotes the fast time. The transit time $T_f$ of light in the loop is much longer than $T_p$ and defines a slow time $t$ (round-trip number) in such a way that the physical time $\tau$ is given by $\tau=tT_f+k$, with $t=0,1,2,3,..$ and $0<k<T_f$ \cite{r81}. A phase modulator impresses a periodic phase change $\Delta \theta_t(k)= H(k)+ \xi_t P(k)$ to the optical pulse after each transit $t$ in the loop, where the fast time $k$ is given in units of the modulation period $T_m$ and $\xi_t$ are independent Gaussian variables with zero mean and unit variance, i.e. 
$\overline{\xi_t}=0$ and $\overline{ \xi_t \xi_{t^{\prime}}}= \delta_{t, t^{\prime}}$. An optical amplifier is placed inside the loop to compensate for the coupler losses. In this way, the envelope $\Psi(k,t)$ of the optical pulse at the $t$-th transit in the loop reads 
\begin{equation}
\Psi(k,t)=\Psi(k,0) \exp[-i H(k)t-i P(k) W(t)]
\end{equation}
where $W(t)=\sum_{l=0}^{t} \xi_l$ is a stroboscopic map of a Wiener process at discrete times $t$, i.e. $\overline{W(t)}=0$ and $\overline{W^2(t)}=t$. A comparison of Eqs.(6) and (9) shows that the pulse evolution at successive transits in the loop describes a stroboscopic map of a quantum trajectory in Bloch space of a dissipative quantum system with effective non-Hermitian Hamiltonian $H_{eff}(k)=H(k)-(i/2) P^2(k)$. The spectral content (frequency comb) $\psi_n(t)= \int dk \Psi(k,t) \exp(ikn)$ of the optical pulse at successive transits, which provides the site occupation amplitudes $\langle n | \psi(t) \rangle$ of the synthetic lattice, can be monitored at the output port of the coupler. The analogue of the density matrix $\rho_{n,m}$ is the noise-averaged spectral matrix $\overline {\psi_n \psi_m^*}$, i.e. so-called mutual coherence or complex visibility \cite{r79,r80}, which  can be measured by suitable techniques \cite{r80}. The short-time relaxation dynamics of density matrix, with the characteristic elongated pattern for $\varphi \neq 0$ shown in Fig.2, could be thus accessed in the optical platform.\\
\\
{\it Conclusion.} In this work we reconsidered a paradigmatic effect of non-Bloch band theory, the non-Hermitian skin-effect, in the framework of open quantum systems.
Our results reveal hidden signatures of the skin effect beyond the non-Hermitian theory, suggesting that quantities used in quantum statistical mechanics, like von Neumann entropy, might provide a pivotal role in characterizing dissipative topological matter beyond semiclassical models. Noise-driven classical systems, such as stochastic-driven spectral photonic lattices, could provide a suitable platform for the observation of jump dynamics and hidden signatures of the skin effect.\\
\\
The author acknowledges
the Spanish State Research Agency through the Severo Ochoa
and Maria de Maetzu Program for Centers and Units of Excellence
in R\&D (Grant No. MDM-2017-0711).
%\clearpage
%\bibliography{H:/Physik/bibliography}

\newpage
\begin{widetext}
%\usepackage[format=plain,labelfont=bf,up,textfont=normal,up,figurename=Fig.~S]{caption}
%\captionsetup[figure]{labelfont=bf,textfont=normalfont,singlelinecheck=off,justification=raggedright}

\begin{center}
\section*{ \bf Supplemental Material}% for \\$^{\prime}$Topological phase transition in non-Hermitian quasicrystals$^{\prime}$}
\end{center}
\renewcommand{\thesubsection}{S}
\renewcommand{\theequation}{S-\arabic{equation}}
\setcounter{equation}{0}
\begin{center}
{\it {\bf S.1. Stochastic Schr\"odinger equation and Lindblad master equation: collective jump operator model and mean-field limit}}\par
\end{center}
The single-particle Schr\"odinger equation with a non-Hermitian Hamiltonian $\hat{H}_{eff}=\hat{H}- (i/2) \hat{P}^2$, i.e.
\begin{equation}
i \partial_t | \psi(t) \rangle= \hat{H}_{eff} | \psi(t) \rangle ,
\end{equation}
describes a non-unitary dynamics since the norm of the state vector $|\psi(t) \rangle$ is not conserved. To restore a unitary dynamics, suitable noise terms should be added to the right hand side of Eq.(S-1), resulting in a
 Schr\"odinger-Langevin equation [1]
\begin{equation}
i d |\psi(t) \rangle= dt \hat{H}_{eff} | \psi(t) \rangle +\sum_l \hat{\Gamma}_l | \psi(t) \rangle \xi_l(t)dt
\end{equation}
where $\hat{\Gamma}_l$ is a set of operators acting on the stochastic wave function $| \psi(t) \rangle$ and $\xi_l(t)$ are zero-mean delta-correlated Gaussian white noise terms 
\begin{equation}
\overline{ \xi_l(t)}=0 \; ,\;\; \overline{\xi^*_n(t) \xi_l(t^{\prime})}= \delta_{n,l} \delta (t - t^{\prime})
\end{equation}
(the overline denotes ensemble average).
Equation (S-2) is written here using the It\^{o} interpretation of stochastic differential equations [1,2]. Note that $dW_n(dt)=\int_{t}^{t+dt} dt^{\prime} \xi_n(t^{\prime})$ describe Wiener processes 
with zero mean 
\begin{equation}
\overline{dW_n(dt)}=0
\end{equation}
and correlations
\begin{equation}
\overline {dW_n^*(dt) dW_l (dt^{\prime})}=\delta_{n,l} {\rm min} (dt, dt^{\prime}).
\end{equation}
Hence quadratic terms in $dW_n$, which scale as $ \sim dt$, should be included in differential calculus. The choice of the operators $\hat{\Gamma_l}$ in Eq.(S-2) that gives (on average) a unitary evolution is not unique and basically depends on the microscopic details of the dissipative system-bath coupling. From Eq.(S-2), at leading order in $dt$ the norm at time $t+dt$ can be readily calculated as
\begin{eqnarray}
\langle \psi(t+dt) | \psi(t+dt) \rangle & = & \langle \psi(t) | \psi(t) \rangle-\langle \psi(t) | \hat{P}^2 \psi(t) \rangle dt 
-i \sum_l dW_l(dt) \langle \psi(t) | \hat{\Gamma}_l \psi(t) \rangle +i \sum_l dW_l^*(dt) \langle \psi(t) | \hat{\Gamma}_l^{\dag} \psi(t) \rangle \nonumber \\
& + & \sum_{n,l}dW_l^*(dt)dW_n(dt) \langle \psi(t) | \hat{\Gamma}_l^{\dag} \hat{\Gamma}_n \psi(t) \rangle . 
\end{eqnarray}
 Using Eqs.(S-4) and (S-5), after ensemble average one obtains
\begin{equation}
{ \langle \psi(t+dt) | \psi(t+dt) \rangle } = { \langle \psi(t) | \psi(t) \rangle}
+ \langle \psi(t) |  \left( \sum_{n} \hat{\Gamma}_n^{\dag} \hat{\Gamma}_n -\hat{P}^2 \right) \psi(t) \rangle  dt 
\end{equation}
 Hence the only condition that has to be satisfied to obtain a norm-preserving dynamics is
 \begin{equation}
 \hat{P}^2= \sum_n \hat{\Gamma}_n^{\dag}\hat{\Gamma}_n.
 \end{equation}
 which provides the analogue of the fluctuation-dissipation theorem for classical systems [1]. The {\it minimal} model that realizes a unitary dynamics is obtained by assuming a single stochastic term in Eq.(S-2), namely $\hat{\Gamma_l}=\hat{\Gamma}_l^{\dag}=\hat{P}$.
 In this case the  Schr\"odinger-Langevin equation (S-2) takes the form 
 \begin{equation}
 i d | \psi(t) \rangle = dt \hat{H}_{eff} | \psi(t) \rangle+ dW \hat{P} | \psi(t) \rangle \;\;\;\; {\rm( Ito)} 
 \end{equation}
 which is precisely Eq.(4) given in the main text. 
  Note that the Stratonovich form of the stochastic Schr\"odinger equation reads
 \begin{equation}
 i d | \psi(t) \rangle = dt \hat{H}| \psi(t) \rangle+ dW \hat{P} | \psi(t) \rangle \;\;\;\; {\rm( Stratonovich)}
 \end{equation}
 which is obtained from Eq.(S-9) after the replacement $\hat{H}_{eff} \rightarrow \hat{H}$. The temporal evolution of the {\it ensemble mean value} of the state vector, $ \overline{| \psi(t) \rangle}$, can be readily obtained from the It\^{o} form after taking the ensemble average of both sides of Eq.(S-9). Since in the It\^{o} calculus $|\psi(t) \rangle$ is independent of $dW$  and $\overline{dW}=0$ [2], one has $\overline {| \psi (t) \rangle dW}=0$ and thus
 \begin{equation}
 i \frac{d}{dt} \overline{| \psi(t) \rangle}= \hat{H}_{eff} \overline{| \psi(t) \rangle},
 \end{equation}
  i.e. the mean wave function $\overline{| \psi(t) \rangle}$ satisfies the deterministic Schr\"odinger equation with the effective non-Hermitian Hamiltonian $ \hat{H}_{eff}$. This corresponds to the so-called mean-field (or semiclassical) limit of the Schr\"odinger-Langevin equation, where the stochastic (jump) term in Eq.(S-9) is neglected.\\ 
 The density operator $\hat{\rho}(t)= | \psi(t) \rangle \langle \psi(t)|$ associated to the stochastic Schr\"odinger equation satisfies a master equation of Lindblad form with one collective jump operator given by $\hat{P}$. In fact, one has
 \begin{equation}
 \hat{\rho}(t+dt)= \hat{U}(dt) \hat{\rho}(t) \hat{U}^{\dag}(dt)
 \end{equation}
 where 
\begin{equation} 
 \hat{U}(dt)=\exp(-i \hat{H}dt-i \hat{P}dW)=1-i \hat{H}dt -i \hat{P}dW-\frac{1}{2}\hat{P}^2 dW^2
 \end{equation}
is the propagator of the Schr\"odinger-Langevin equation for an infinitesimal time step $dt$. Substitution of Eq.(S-13) into Eq.(S-12), after averaging over $dW$ one obtains
\begin{equation}
 \hat{\rho}(t+dt)= \hat{\rho}(t)-i [ \hat{H} dt, \hat{\rho}(t)]- \frac{1}{2} \hat{P}^2 \hat{\rho}(t)dt - \frac{1}{2} \hat{\rho}(t) \hat{P}^2dt +\hat{P} \hat{\rho}(t) \hat{P}dt.
\end{equation}
 which corresponds to the Lindblad master equation given in the main text [Eq.(5)]. The semiclassical limit in the Lindblad master equation [3] corresponds to neglect the last term on the right side of Eq.(S-14), i.e. the quantum jump term. \\
 \begin{center}
{\it {\bf S.2. Relaxation dynamics}}\par
\end{center}
{\it Short-time (bulk) relaxation dynamics.} Let us consider a lattice in the thermodynamic limit $N \rightarrow \infty$, so that edge effects can be neglected. In this case the stochastic Schr\"odinger equation (S-9) can be solved in momentum space, where all operators are diagonal. After expanding the state vector $| \psi(t) \rangle$ in Bloch basis as
\begin{equation}
 |\psi(t) \rangle = \int_{-\pi}^{\pi}  dk \Psi(k,t) | k \rangle,
 \end{equation}
  where $k$ is the Bloch wave number (which varies in the range $-\pi \leq k < \pi$) and $| k \rangle \equiv(1/ \sqrt{2 \pi}) \sum_n \exp(ikn) | n \rangle$ is the Bloch basis, from Eq.(S-9) one obtains
  \begin{equation}
 i d \Psi(k,t)  =  \left[ H(k)-\frac{i}{2} P^2(k) \right] dt \Psi(k,t)+dW P(k) \Psi(k,t),
  \end{equation}
  i.e.
  \begin{equation}
  \Psi(k,t+dt)=\exp \left\{ -i H(k) dt -i P(k) dW \right\} \Psi(k,t).
  \end{equation}
  Equation (S-17) can be iterated, yielding 
  \begin{equation}
  \Psi(k,t)=\exp \left\{ -i H(k)t -i P(k) W(t)  \right\} \Psi(k,0)
  \end{equation}  
  where $W(t)=\int_0^t dt^{\prime} \xi(t^{\prime})$ is a Winer process of zero mean and variance $t$, i.e. $\overline{W}=0$ and $\overline{W^2}=t$. The density operator $\hat{\rho}(t)$ in Bloch basis is readily obtained as
  \begin{equation}
  \rho_{k, k^{\prime}}(t) \equiv \langle k | \hat{\rho} (t) |  k^{\prime} \rangle = \overline{\Psi(k,t) \Psi^*(k^{\prime},t)}
  \end{equation}
  where the overbar denotes the ensemble average over noise realizations. Taking into account that for any scalar $f$ the following identity holds [2]
  \begin{equation}
  \overline{\exp(-i f W)}= \exp  (-f^2 \overline{W^2} /2)=\exp  (-f^2 t/2)
  \end{equation}
  from Eqs.(S-18), and (S-19) and (S-20) one obtains
  \begin{equation}
  \rho_{k, k^{\prime}}(t)  =  \rho_{k, k^{\prime}}(0) \exp [iH(k^{\prime})t-iH(k) t]   \times \exp \left\{ -\frac{1}{2}[P(k^{\prime})-P(k)]^2 t \right\}
  \end{equation}
  which is precisely Eq.(7) given in the main text. The matrix elements of the density operator $\hat{\rho}(t)$ in the Wannier basis, $\rho_{n,m}(t)= \langle n | \hat{\rho}(t) | m \rangle$, are then 
 computed from the relation
 \begin{equation}
 \rho_{n,m}(t)= \iint dk dk^{\prime} \rho_{k,k^{\prime}}(t)  \langle n | k \rangle \langle k^{\prime}   | m \rangle
 \end{equation} 
  i.e.
 \begin{equation}
 \rho_{n,m}(t)= \frac{1}{2 \pi} \iint dk dk^{\prime} \rho_{k,k^{\prime}} (t) \exp(ikn-ik^{\prime}m)
  \end{equation} 
  where the integrals are extended over the square domain $-\pi \leq k < \pi$, $ -\pi \leq k^{\prime} < \pi$.
  Substitution of Eq.(S-21) into Eq.(S-23) and taking into account that
  \begin{equation}
  \rho_{k,k^{\prime}} (0)=\frac{1}{2 \pi} \sum_{l,q} \rho_{l,q}(0) \exp(ik^{\prime}q-ikl)
  \end{equation}
  one obtains
  \begin{equation}
 \rho_{n,m}(t)= \frac{1}{4 \pi^2}  \sum_{l,q} \rho_{l,q}(0) \iint dk dk^{\prime}  \exp \left\{ ik(n-l)-ik^{\prime}(m-q) +iH(k^{\prime})t-iH(k)t  -\frac{1}{2}[P(k^{\prime})-P(k)]^2 t \right\}  
 \end{equation} 
  In particular, if the system is initially prepared in a pure state with the excitation localized at the single site $n=0$, i.e. $\rho_{l,q}(0)=\delta_{l,0} \delta_{q,0}$, one has
  \begin{equation}
   \rho_{n,m}(t)= \frac{1}{4 \pi^2}  \iint dk dk^{\prime} \exp \left\{ ikn-ik^{\prime}m +iH(k^{\prime})t-iH(k)t  -\frac{1}{2}[P(k^{\prime})-P(k)]^2 t \right\}
   \end{equation}
  which coincides with Eq.(8) given in the main text. We note that, if $P(-k)=P(k)$ the ballistic spreading of $\rho_{n,m}(t)$ as described by Eq.(S-26) is highly symmetric because the following conditions
  \begin{equation}
  \rho_{n,-m}(t)=\rho_{-n,m}(t)=\rho_{-n,-m}(t)=\rho_{n,m}(t)
  \end{equation}
  hold at any time, in addition to the Hermiticity constraint $\rho_{n,m}(t)=\rho^*_{m,n}(t)$ of the density matrix. Conversely, when $P(-k) \neq P(k)$ most of such symmetries are lost. This results in qualitatively distinct spreading patterns, as shown in Fig.2 of the main manuscript. Remarkably, for $P(k) \neq P(-k)$ the slower decaying elements of $\rho_{n,m}(t)$ lie on the diagonal $n=m$, corresponding to populations, while for $P(k)=P(-k)$ also the elements on the anti-diagonal $m=-n$, corresponding to coherences, are slowly decaying [owing to the symmetry constraint Eq.(S-27)]. To prove this statement, let us consider the 
 asymptotic behavior of $\rho_{n,m}(t)$ for sufficiently long times $t$ (yet short enough to avoid edge effects) along the $^{\prime}$line$^{\prime}$ $n= \alpha t$ and $m=\beta t$ of the $(n,m)$ plane, with $\alpha$ and $\beta$ real parameters which are left undetermined at this stage. Note that for $\alpha=\beta$ the line scans the diagonal elements (populations) of the density matrix ($n=m$), while for $\alpha=-\beta$ the line scans the anti-diagonal elements $m=-n$ corresponding to coherences.  
 Equation (S-26) entails to calculate the double integral
 \begin{equation}
 I(t)=\iint dk dk^{\prime} \exp \left\{ - F(k,k^{\prime}) t  \right\}
 \end{equation}
  at long times $t$, where we have set
  \begin{equation}
  F(k,k^{\prime}) \equiv \frac{1}{2} [P(k)-P(k^{\prime})]^2-i \alpha k+i \beta k^{\prime}-iH( k^{\prime})+iH(k).
  \end{equation}
  Clearly, at long times the main contribution to the integral $I(t)$ comes from the line $\mathcal{C}$ of the $(k,k^{\prime})$ plane implicitly defined by the cartesian equation
  \begin{equation}
  P(k)=P(k^{\prime}).
  \end{equation}  
Correspondingly, the exponential decaying term under the sign of the integral on the right hand side of Eq.(S-28), determined by the real part of $F$, vanishes. Moreover, according to the multivariate saddle point criterion of asymptotic analysis [4], $I(t)$ decays at least like $ \sim 1/t$ whenever there are saddle points $(k,k^{\prime})$ of $F$ on $\mathcal{C}$. These are obtained from the condition ${\rm grad} F=0$, which implies
\begin{equation}
\alpha=H^{\prime}(k)\; , \; \; \; \beta=H^{\prime}(k^{\prime}).
\end{equation}
Equation (S-31) defines the direction in the $(n,m)$ plane and corresponding ballistic spreading velocities where the density matrix elements $\rho_{n,m}(t)$ decay like $\sim 1 / t$ once Eq.(S-30) is satisfied for some $(k, k^{\prime})$. The decay is even slower than $ \sim 1 /t$ provided that, in addition to ${\rm grad} F=0$, also the Hessian of the form $F(k,k^{\prime})$ vanishes, i.e.
\begin{equation}
\left|  
\begin{array}{cc}
 \frac{\partial^2 F}{\partial k^2} & \frac{\partial^2 F}{\partial k \partial k^{\prime}} \\
 \frac{\partial^2 F}{ \partial k^{\prime} \partial k}  & \frac{\partial^2 F}{ \partial k^{\prime 2} }
 \end{array}
 \right|=0
\end{equation}
which reads explicitly
\begin{eqnarray}
H^{''}(k) H^{''} (k^{\prime}) & = & 0 \\
H^{''}(k) (P^{\prime}(k^{\prime}))^2 & = & H^{''}(k^{\prime}) (P^{\prime}(k))^2.
\end{eqnarray}
Clearly, the line $\mathcal{C}$ contains the bisector $k=k^{\prime}$, along which conditions (S-33) and (S-34) are identically satisfied at the inflection of $H(k)$. Since for $k=k^{\prime}$ Eq.(S-31) implies $\alpha=\beta$, i.e. $n=m$, we can conclude that the terms of the density matrix $\rho_{n,m}(t)$ along the main diagonal $n=m$, i.e. populations, decay slower than $\sim 1/t$.  The population spreading velocity, $\alpha=H^{\prime}(k)$, is the largest group velocity in the lattice attained at the inflection points $k=k_0$ where $H^{??}(k_0)=0$. Note that, since $H(-k)=H(k)$, the inflection points always appear in pairs, $k_0$ and $-k_0$, with opposite group velocities $H^{\prime}(k_0)=-H^{\prime}(-k_0)$. This means the population spreadings is {\em always} bidirectional. In Fig.2 of the main manuscript, the slowly decaying terms of populations spreading in opposite directions are visible as the brightest spots on the main diagonal $m=n$ that propagate with a velocity $v \sim H^{\prime}(k_0)= \alpha$. Let us now focus our attention to the decay of the coherences $\rho_{n,m}(t)$ on the main anti-diagonal $m=-n$, which  corresponds to the choice $ \alpha=-\beta$ and thus $k^{\prime}=-k$ according to Eq.(S-31) [note that, since $H(-k)=H(k)$, one has $H^{\prime}(-k)=-H^{\prime}(k)$]. If $P(-k)=P(k)$, i.e. if the semiclassical dynamics does not show the NHSE, for the symmetry constraint (S-27) it follows that also the coherences $\rho_{n,-n}(t)$ display the same decay as populations $\rho_{n,n}(t)$, i.e. slower than $\sim 1 /t$. Indeed, Eqs.(S-30), (S-33) and (S-34) are satisfied for $k^{\prime}=-k$ as well. On the other hand, if $P(-k) \neq P(k)$, the bisector $k^{\prime}= -k$ does not belong to the curve $\mathcal{C}$, and thus the decay is faster than $ \sim 1 /t$. Such an asymptotic analysis qualitatively explains the different spreading patterns observed in Fig.2 of the main manuscript for $\varphi=0$ [$P(-k)=P(k)$] and $\varphi \neq 0$  [$P(-k) \neq P(k)$].\\
\\
{\it Long-time relaxation dynamics.} The above analysis is valid provided that the excitation remains  confined far from the edges of the lattice. For a finite lattice made of $N$ sites with open boundaries, the long-time relaxation dynamics is dominated by the non-decaying eigenvectors of the Liouvillian superoperator $\mathcal{L}$ [5-7], i.e. the eigenvectors with a vanishing real part of their eigenvalues. Let us indicate by $\lambda_l$ the $N^2$ eigenvalues of the $N^2 \times N^2$ matrix associated to the Liouvillian superoperator $\mathcal{L}$. We assume that $\mathcal{L}$ is diagonalizable, i.e. we exclude from our analysis the special case where $\mathcal{L}$ shows exceptional points [3]. Since $\mathcal{L}$ is trace preserving, all eigenvalues $\lambda_l$ have a non-positive real part, ${\rm Re}(\lambda_l) \leq 0$, with at least one
zero eigenvalue $\lambda_l=0$, corresponding to a stationary state  $\hat{\rho}^{(s)}$  of $\mathcal{L}$, $\mathcal{L} \hat{\rho}^{(s)}=0 $. For the Lindblad master equation (S-14) it can be readily shown by a direct inspection that the diagonal state
\begin{equation}
\hat{\rho}^{(s)}= \frac{1}{N}
\left(
\begin{array}{cccccc}
1 & 0 & 0 & ... & 0 & 0 \\
0 & 1 & 0 & ... & 0 & 0 \\
0 & 0 & 1 & ... & 0 & 0 \\
... & ... & ... & ... & ... & ... \\
0 & 0 & 0 & ... & 0 & 1
\end{array}
\right),
\end{equation}
 written in Wannier basis, $ \rho_{n,m}= \langle n | \hat{\rho}| m\rangle$, is a stationary state of $\mathcal{L}$. In fact $\hat{\rho}^{(s)}$ commutes with both $\hat{H}$ and $\hat{P}$, and thus $\mathcal{L}\hat{\rho}^{(s)}=0$. Such a stationary state corresponds to a maximally mixed state, i.e. the von Neumann entropy 
 \begin{equation}
 S(\hat{\rho})=-{\rm tr} ( \rho \log \rho)
 \end{equation}
for this state reaches the largest value $S(\hat{\rho}^{(s)})= \ln N$.\\ 
Interestingly, as there is always one stationary state, its uniqueness is not guaranteed [5-7].  In particular, the existence of certain symmetries gives rise to multiple stationary states [5,7]. In our case, $\hat{H}$ is assumed to satisfy the condition $\hat{H}^T=\hat{H}$ in Wannier basis ($T$ is the transpose operation), i.e. $H(-k)=H(k)$ in Bloch basis, because of time reversal symmetry. Let us now assume that $\hat{H}_{eff}$ does not show the NHSE. This is equivalent to state that also $\hat{P}$ is invariant, in Wannier basis, under the transpose operation, or likewise $P(-k)=P(k)$ in Bloch basis. In this case, it can be shown that invariance by transposition for {\it both} $\hat{H}$ and $\hat{P}$  ensures that the bi-diagonal state
\begin{equation}
\langle n | \hat{\rho}^{(s_a)} | m \rangle = \frac{1}{N+1} \left( \delta_{n,m}+\delta_{m,N-n+1} \right),
\end{equation}
is a stationary state of $\mathcal{L}$, satisfying the constraints $\hat{\rho}^{\dag}=\hat{\rho}$,  ${\rm Tr}(\hat{\rho})=1$ and ${\rm Tr}(\hat{\rho}^2) =2/(N+1)< 1$ required for a density operator. This follows from the fact that, as one can directly prove, $[\hat{H},\hat{\rho}^{(s_a)}]=[\hat{P}, \hat{\rho}^{(s_a)}]=0$ and thus $\mathcal{L} \rho^{(s_a)}=0$. Therefore, {\it if in the semiclassical limit the dissipative dynamics described by $\hat{H}_{eff}$ does not show the NHSE, there are at least two stationary states of the Lindblad master equation, and the final state after the relaxation process depends rather generally on the initial state.} This means that, for a rather arbitrary initial state $\hat{\rho}(0)$, the asymptotic state $\hat{\rho}(\infty)$ after the relaxation process does not maximize the von-Neumann entropy $S$ (see the curve in the inset of Fig.3 of the main text corresponding to the phase $\varphi=0$).\\ 
Finally, we note that a larger number of stationary states can be found in case of lattices with nearest-neighbor hopping, like in the model described by Eq.(3) of the main text with $T=0$. In this case the matrices associated to $\hat{H}$  
and $\hat{P}$ are tridiagonal, and under the conditions $\hat{H}^T=\hat{H}$ and  $\hat{P}^T=\hat{P}$, they are Jacobi matrices which commute, i.e. $[ \hat{H},\hat{P}]=0$. In this case, $\hat{H}$ and $\hat{P}$ share the same set of eigenfunctions $ | \phi^{(\alpha)} \rangle$, which read explicitly 
\begin{equation}
| \phi^{(\alpha)} \rangle=  \sqrt{\frac{2}{ N+1}} \sum_{n=1}^{N} \sin \left( \frac{\pi n \alpha}{N+1} \right) | n \rangle
\end{equation}
($\alpha=1,2,...,N$). The all set of eigenvalues $\lambda_l$ and corresponding eigenvectors $\rho^{(l)}$ of the Liouvillian superoperators can be readily computed in a closed form as follows. Let 
\begin{equation}
\hat{H} | \phi^{(\alpha)} \rangle=E_{\alpha} | \phi^{(\alpha)} \rangle \; , \; \; \;  \hat{P} | \phi^{(\alpha)} \rangle=p_{\alpha} | \phi^{(\alpha)} \rangle.
\end{equation}
Then the state
\begin{equation}
\rho^{(l)}=  | \phi^{(\alpha)} \rangle  \langle \phi^{(\beta)} | 
\end{equation}
($\alpha, \beta=1,2,...,N$, $l=\alpha+N(\beta-1)=1,2,...,N^2$) is an eigenfunction of $\mathcal{L}$ with eigenvalue 
\begin{equation}
\lambda_l=i(E_{\beta}-E_{\alpha})-\frac{1}{2} (p_{\alpha}-p_{\beta})^2,
\end{equation}
i.e. $\mathcal{L} \hat{\rho}^{(l)}= \lambda_l  \hat{\rho}^{(l)}$. Since for $\alpha=\beta$ one has $\lambda_l=0$, it follows that there are exactly $N$ stationary states of $\mathcal{L}$ with eigenfunctions $| \phi^{(\alpha)} \rangle  \langle \phi^{(\alpha)} | $ ($\alpha=1,2,...,N$). Note that in this degenerate case the two stationary states $\hat{\rho}^{(s)}$ and $\hat{\rho}^{(sa)}$, given by Eqs.(S-35) and (S-37), are just linear combinations of  states 
$| \phi^{(\alpha)} \rangle  \langle \phi^{(\alpha)} | $.
\\
 \begin{figure*}
\includegraphics[width=16cm]{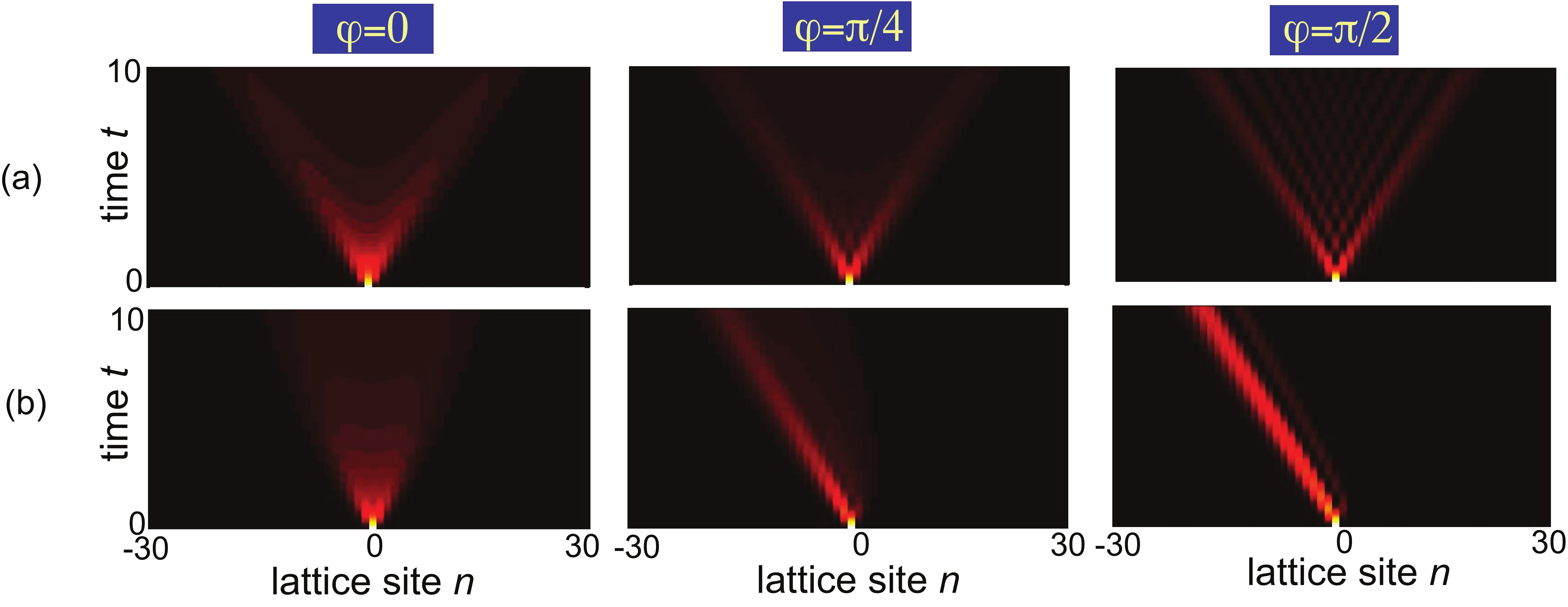}
\caption{(Color online) Short-time (bulk) evolution of the diagonal density matrix elements (snapshots of $|\rho_{n,m}(t)|$ on a pseuodocolor map) in the dissipative lattice defined by Eq.(3) for $J=R=1$, $T=0$ and for three values of the phase $\varphi$ ($0$, $ \pi/4$ and $\pi/2$). The initial state is the pure state $\rho (0) = |0 \rangle \langle 0|$, corresponding to a particle localized at site $n=0$. Panels (a) refer to the predictions based on the master equation (5) with a  collective jump operator, while panels (b) are the predictions based on the non-Hermitian Hamiltonian ${H}_{eff}$ without jumps (semiclassical limit). Note that, for a non-vanishing phase $\varphi$, in (b) the population drifts along the lattice with some drift velocity $v$, according to the general results of Ref.[9].}
\end{figure*}

 {\it Non-Hermitian unidirectional flow in the semiclassical limit.} The short-time (bulk) relaxation dynamics of the diagonal elements $\rho_{n,n}(t)$ of density operator (populations) for the stochastic Schr\"odinger equation, depicted in Fig.2 of the main manuscript, shows a symmetric ballistic spreading from the initially excited site, regardless of the value of the phase $\varphi$. This result is is sharp contrast with the predictions of the semiclassical limit, i.e. when disregarding the jump term and considering the population evolution for the non-Hermitian Hamiltonian ${H}_{eff}$. In this case, for a non-vanishing phase $\varphi$ the NHSE is responsible for a unidirectional flow of the population along the lattice with some drift velocity, which can be determined from a saddle point analysis [9]. Such a distinct behavior is illustrated in Fig.6, which compares the spreading dynamics of the populations $\rho_{n,n}(t)$ for the full master equation (5) [panels (a)] and in the semiclassical limit [panels (b)].

 \begin{center} 
{\it {\bf S.3 Dissipative Hatano-Nelson model}}\par
\end{center}
A paradigmatic non-Hermitian model displaying the NHSE is provided by the Hatano-Nelson model [8], which describes particle hopping on a tight-binding lattice with asymmetric left/right hopping $J_1$ and $J_2>J_1$. 
The Schr\"odinger equation in Wannier basis reads
\begin{equation}
i \frac{d \psi_n}{dt}= J_1 \psi_{n+1}+J_2 \psi_{n-1} -i(J_2-J_1) \psi_n \equiv \sum_l (H_{eff})_{n,l} \psi_l
\end{equation}
where the on-site dissipative term on the right hand side of Eq.(S-42) has been added to ensure that the system is purely dissipative (all eigenvalues of $H_{eff}$ have non-positive imaginary part). Under PBC, the Bloch energy spectrum reads
\begin{equation}
E(k)=(J_1+J_2) \cos (k)-i(J_2-J_1) [1+ \sin (k)]
\end{equation}
  \begin{figure*}
\includegraphics[width=17.6cm]{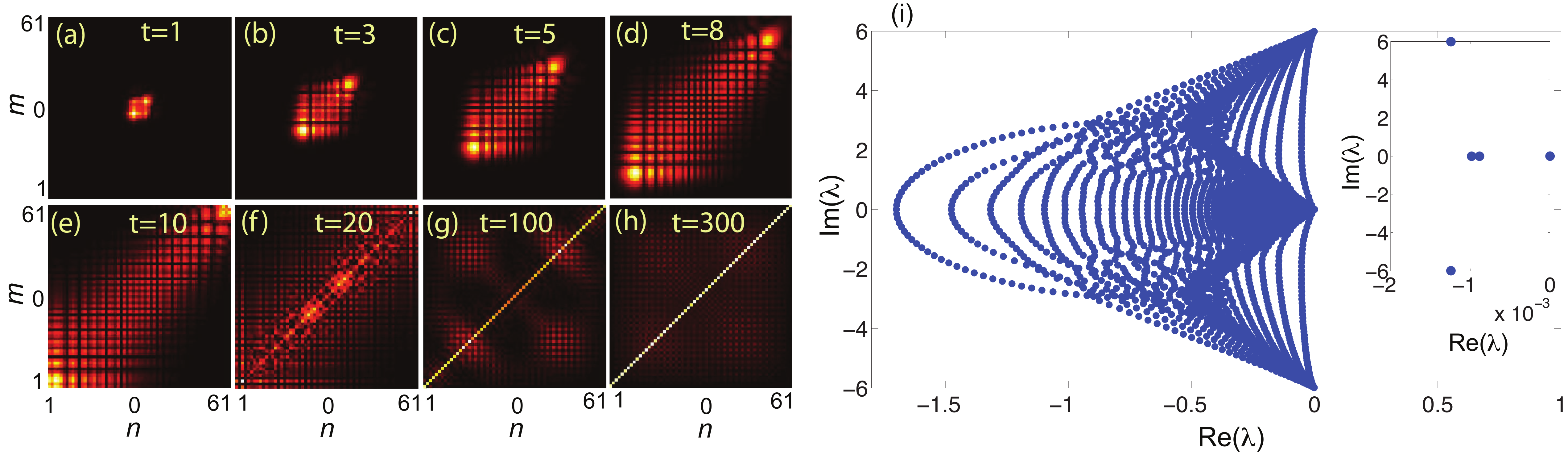}
\caption{(Color online) (a-h) Evolution of the density matrix (snapshots of $|\rho_{n,m}(t)|$ on a pseuodocolor map) in the dissipative Hatano-Nelson lattice with open boundaries for parameter values $J_1=1$, $J_2=2$, and $N=61$. The initial condition is the pure state $\hat{\rho}(0)=|n \rangle \langle n|$ with $n=(N+1)/2=31$. (i) Numerically-computed $N^2$ eigenvalues $\lambda_l$ of the Liouvillian superoperator $\mathcal{L}$. The inset in the figure shows an enlargement of the eigenvalues near $\lambda=0$.}
\end{figure*}

which describes an ellipse in complex energy plane. However, under OBC the energy spectrum collapses to a segment in the interior of the ellipse, which is associated to the NHSE.
For a lattice comprising $N$ sites under OBC, the effective non-Hermitian matrix Hamiltonian $H_{eff}$ can be written as $H_{eff}={H}-(i/2)P^2$, where $H$ and $P^2$ are $N \times N$ tridiagonal Hermitian matrices given by
\begin{equation}
H= \left(
\begin{array}{ccccccc}
0 & \frac{J_1+J_2}{2} & 0 & ... & 0 & 0 & 0 \\
\frac{J_1+J_2}{2} & 0 & \frac{J_1+J_2}{2} & ... & 0 & 0  & 0 \\
0 & \frac{J_1+J_2}{2} & 0 & ... & 0 & 0  & 0  \\
... & ... & ... & ... & ... & ... & ... \\
0 & 0 & 0 & ... & \frac{J_1+J_2}{2} & 0 & \frac{J_1+J_2}{2} \\
0 & 0 & 0 & ... & 0 & \frac{J_1+J_2}{2} & 0
\end{array}
\right)
\end{equation}
\begin{equation}
P^2= \left(
\begin{array}{ccccccc}
2(J_2-J_1) &-i(J_2-J_1) & 0 & ... & 0 & 0 & 0 \\
i(J_2-J_1) & 2 (J_2-J_1) & -i(J_2-J_1) & ... & 0 & 0  & 0 \\
0 & i(J_2-J_1) & 2(J_2-J_1) & ... & 0 & 0  & 0  \\
... & ... & ... & ... & ... & ... & ... \\
0 & 0 & 0 & ... & i(J_2-J_1) & 2(J_2-J_1) & -i(J_2-J_1)\\
0 & 0 & 0 & ... & 0 & i(J_2-J_1) & 2(J_2-J_1)
\end{array}
\right).
\end{equation}
To calculate the jump operator $P$ entering in the master equation, i.e. the square root of the matrix $P^2$, let us notice that $\hat{P}^2= \sum_{n,m} (P^2)_{n,m} |n \rangle \langle m| $ can be readily diagonalized as 
\begin{equation}
\hat{P}^2= \sum_{\alpha=1}^{N} E_{\alpha} | \alpha \rangle \langle \alpha |
 \end{equation}
 where the eigenvalues $E_{\alpha}$ and corresponding eigenvectors $| \alpha \rangle$ of $\hat{P}^2$ read explicitly
 \begin{equation}
 E_{\alpha}=2 (J_2-J_1) \left[ 1 + \cos\left( \frac{ \pi \alpha}{N+1} \right) \right]
 \end{equation}
 \begin{equation}
 \langle n | \alpha \rangle=\sqrt{\frac{2}{N+1}} \sin \left( \frac{\pi \alpha n}{N+1}  \right) \exp(i \pi n/2)
 \end{equation}
 ($ \alpha=1,2,3,...,N$). The operator $\hat{P}$ is then given by ${\hat P}=\sum_{\alpha} \sqrt{E_{\alpha}} | \alpha \rangle \langle \alpha |$, i.e. in Wannier basis ${\hat P}= \sum_{n,m} P_{n,m} |n \rangle \langle m|$ where
 \begin{equation}
 P_{n,m}= \frac{2}{N+1} \sum_{\alpha=1}^N \ \sqrt{2 (J_2-J_1) \left[ 1 + \cos\left( \frac{ \pi \alpha}{N+1} \right) \right]} \sin \left( \frac{\pi n \alpha}{N+1} \right) \sin \left( \frac{\pi m \alpha}{N+1} \right) \exp[i \pi(n-m)/2] 
 \end{equation}
 are the elements of the matrix $P$.
Note that, while $P^2$ is short-range (tridiagonal) matrix, the jump matrix $P$ in the minimal model of the dissipative Hatano-Nelson Hamiltonian is a long-range matrix. 
Figure 7 shows, as an example, the evolution of the density matrix $\rho_{n,m}(t)$ in Wannier basis in a lattice with OBC comprising $N=61$ sites, as obtained by numerical solving the master equation (S-14) with the initial condition $\hat{\rho}(0)= | n \rangle \langle n|$ with $n=31$, corresponding to a pure state with the particle placed at the $n$-th site of the lattice. The hopping amplitudes are $J_1=1$ and $J_2=2$. The short-time relaxation dynamics, i.e. before the excitation reaches the edges of the lattice [panels (a-d) in Fig.7] clearly shows the characteristic ballistic spreading elongated along the main diagonal $m=n$. Note that, contrary to the unidirectional flow of excitation predicted in the semiclassical limit and associated to the NHSE [9] (see Fig.6 discussed above), the spreading in the $^{\prime}$populations$^{\prime}$ $\rho_{n,n}(t)$ (the diagonal elements of the density matrix) remains almost symmetric around $n=0$, so that in the dissipative description of the Hatano-Nelson model beyond the semiclassical limit the unidirectional flow is washed out when looking at the spreading of excitation solely. However, the elongated spreading pattern of the {\em entire} density matrix elements along the main diagonal $n=m$ provides the hidden signature of the NHSE found in the semiclassical limit. In the long time limit, where edge effects become important [panels (e-h) in Fig.7], the density matrix relaxes toward the maximally-mixed stationary state $\hat{\rho}^{(s)}$ [Eq.(S-35)]. The numerically-computed eigenvalues of the Liouvillian superoperator $\mathcal{L}$ is shown in Fig.7(i). The spectrum shows a simple zero eigenvalue, associated to the stationary state $\hat{\rho}^{(s)}$, while all other eigenvalues have a strictly negative real part and correspond to decaying states.\\  
\\

%\newpage 
\small

\noindent
{\bf References}\\
\\
{[1]} N.G. van Kampe, {\it Stochastic Processes in Physics and Chemistry} (Elsevier Science B.V., Amsterdam,  North Holland, 2003).\\
{[2]} C.W. Gardiner, {\it Handbook of Stochastic Methods} (Springer, Berlin, 1985).\\
{[3]} F. Minganti, A. Miranowicz, R.W. Chhajlany, and F. Nori, Quantum exceptional points of non-Hermitian Hamiltonians and
Liouvillians: The effects of quantum jumps, 
Phys. Rev. A {\bf 100}, 062131 (2019).\\
{[4]} T. Neuschel, Ap\'ery Polynomials and the Multivariate Saddle Point
Method, Constr. Approx. {\bf 40}, 487 (2014).\\
{[5]} V.V. Albert and L. Jang, Symmetries and conserved quantities in Lindblad master equations, Phys. Rev. A {\bf 89}, 022118 (2014).\\
{[6]} F. Minganti, A. Biella, N. Bartolo, and C. Ciuti,
Spectral theory of Liouvillians for dissipative phase transitions, Phys. Rev. A {\bf 98}, 042118 (2018).\\
{[7]} D. Manzano and P.I. Hurtado, Harnessing symmetry to control quantum transport, Adv. Phys. {\bf 67}, 1-67 (2018).\\
{[8]} N. Hatano and D. R. Nelson, Localization Transitions
in Non-Hermitian Quantum Mechanics, Phys. Rev. Lett. {\bf 77}, 570 (1996).\\
{[9]} S. Longhi, Probing non-Hermitian skin effect and non-Bloch phase transitions,
Phys. Rev. Research {\bf 1}, 023013 (2019).\\

\end{widetext}
\end{document}